\documentclass[showpacs,amsmath,amsfonts,amssymb,aps,superscriptaddress]{revtex4}
\usepackage{xcolor}
\definecolor{navy}{HTML}{2F729C}
\colorlet{linkcolor}{navy}
\colorlet{headerlinkcolor}{cyan}
\usepackage[colorlinks]{hyperref}[2011/02/05]
\hypersetup{allcolors=linkcolor}
\makeatletter
\newcommand*{\org@sect}{}
\let\org@sect\@sect
\def\@sect#1#2#3#4#5#6[#7]#8{%
  \org@sect{#1}{#2}{#3}{#4}{#5}{#6}[{#7}]{%
    \hypersetup{allcolors=headerlinkcolor}%
    #8%
  }%
}
\makeatother

\usepackage{tabularx}
\usepackage{comment}
\usepackage{enumitem}
\usepackage{graphicx}
\usepackage{float}
\usepackage{dcolumn}
\usepackage{bm}
\usepackage{mathrsfs}
\usepackage{amsmath}

\usepackage{amsfonts}
\usepackage{dsfont}

\begin{document}

\title{Gravitational Collapse via Wheeler-DeWitt Equation}
\author{Davide Batic}
\email{davide.batic@ku.ac.ae}
\affiliation{
Department of Mathematics,\\  Khalifa University of Science and Technology,\\ Main Campus, Abu Dhabi,\\ United Arab Emirates}
\author{M. Nowakowski}
\email{marek.nowakowski@ictp-saifr.org}
\affiliation{
ICTP-South American Institute for Fundamental Research,\\ Rua Dr. Bento Teobaldo Ferraz 271, 
01140-070 S\~ao Paulo, SP Brazil 
}

\date{\today}

\begin{abstract}
We analyze the Wheeler-DeWitt (WDW) equation in the context of a
gravitational collapse. The physics of an expanding/collapsing universe
and many details of a collapsing star can classically be described by
the Roberston-Walker metric in which the WDW equation takes the form
of a times-less Schr\"odinger equation. We set up the corresponding
WDW potential for the collapse and study the solutions of the wave
function.
The results show  that the central singularity appearing in classical
general relativity is avoided, the density is quantized in terms of
the Planck density and the expectation value of the scale factor
exhibits a discrete behavior.
    
\end{abstract}
\pacs{}
\maketitle
\section{Introduction}
Apart from the established semi-classical results like the Hawking
radiation \cite{Hawking} and the Unruh effect \cite{Unruh}, we are
still not sure how the full fledged quantum gravity will look
like. Practically, all candidates ranging from loop quantum gravity
\cite{Loop} over the dynamical triangulation \cite{Triangulation},
causal set theory \cite{Causal}, the path intergral approach
\cite{Path}, non-commutative geometry \cite{Noncommutative} up to
string theory \cite{String}, to mention a few, exhibit certain problems and
obstacles.
An important step towards a theory of quantum gravity has been and
continues to be
till today the WDW equation \cite{WdW} based on the canonical quantization
scheme of gravity. Since its discovery in the sixties up to now, it is
widely used to study quantum phenomena in gravity. Its popularity is
due to the fact that it is a genuine result of a canonical formalism
which makes us supect that in one way or the other it also plays a
role in other theories of quantum gravity. Indeed, it is closely
connected to path integral \cite{Halliwell} and loop quantum
gravity \cite{Smolin1, Smolin2, Mitschul}. In the eighties, the WDW equation was a standard tool
to probe into the early quantum universe \cite{Hartle, Vilenkin, Page,
Kont, Vil} based on the
Robertson-Walker metric. This line of investigation has been carried on
up to now \cite{Braz,Vieira,Gao,He2,Steigl}.
Since the same metric is used for the gravitational collapse
\cite{Weinberg}, in this work, we set up
the WDW equation for this scenario with the hope to get a glimpse how
quantum mechanics affects dense collapsing matter.
The gravitational collapse itself is an active area of inspection and
speculations of different classical and
quantum effects \cite{Bambi, Malafarina}.
Since the
Robertson-Walker metric does not posses any horizon singularity, we can
safely assume that we are indeed examining the quantum spacetime inside
a black hole (BH). Of course, we will not be able to compare our
results with any observation but the expectation of quantum gravity
with respect to the BH is the avoidance of the classical central
singularity.

This approach, as many in the realm of quantum gravity, is not without challenges. The WDW equation 'timeless' nature poses a left out significant hurdle in describing the evolution of a system that is inherently time-dependent in classical terms. But
this 'time-problem'  is an expected outcome of quantum gravity \cite{time1, time2}, and appears also in quantum cosmology. Our study addresses this by interpreting the changes in the wave function, dependent on configuration space variables, as a proxy for dynamical evolution. This method allows us to explore the quantum mechanics of collapsing systems within the existing framework of quantum gravity, despite the absence of a traditional time variable. More precisely, we apply the WDW equation in the late gravitational collapse to study quantum effects in a black hole after all matter has entered the horizon.

The paper is structured as follows: Section I offers an overview
    of the relevant literature on the development of quantum gravity
    theories and their application to astrophysical phenomena. It
    particularly focuses on explaining why the WDW equation can be
    applied to study quantum effects in black holes, specifically
    during the late stage of gravitational collapse after all matter
    has entered the horizon. Section II focuses on the construction of
    the point-like Lagrangian, which plays a fundamental role to the application of  the WDW equation to gravitational collapse scenarios. In Section III, we
    set up the WDW equation specifically for the context of
    gravitational collapse. The results emerging from Section IV
    demonstrate that, in the context of gravitational collapse, the
    central singularity commonly observed in classical general
    relativity is avoided. This is achieved through the quantization
    of the matter density. In Section V, we briefly discuss the issue
    of the time problem. Finally, Section VI offers conclusions and an outlook, reflecting on the implications of our findings and suggesting directions for future research.

\section{Construction of the point-like Lagrangian}
In order to explore the WDW equation in the context
of gravitational collapse, we draw first on analogies from its application in cosmology. We begin by observing that the Friedmann-Robertson-Walker line element with $c=1$ \cite{Weinberg}
\begin{equation}\label{LE}
ds^2=dt^2-R^2(t)\left[\frac{dr^2}{1-kr^2}+r^2 d\vartheta^2+r^2\sin^2{\vartheta}d\varphi^2\right]    
\end{equation}
is valid for both cases. Note that in cosmology, the curvature
parameter $k$ typically takes on values of $\pm 1,0$, rendering $r$ as
dimensionless. Meanwhile, the scale factor $R$ carries a dimension of
length. In the collapse scenario for cold dust, we have \cite{Weinberg}
\begin{equation}
k=\frac{8\pi}{3}G\rho_0,    
\end{equation} 
where $G$ is Newton's gravitational constant and $\rho_0$ is the
density of a spherically symmetric, isotropic and homogeneous dust
cloud. For a representative cloud core having characteristics such as
a radius of $2 \cdot 10^{15}$ Km, a temperature of 10 K, and a mass
that is double the solar mass, we anticipate the initial density,
$\rho_0$, to be approximately $10^{-16}$ Kg/m$^3$ \cite{Bode}.
However, we would not expect that quantum mechanics affects the
collaspe right form the beginning.  It is rather probable that quantum
mechanics sets in after a black hole was formed. Since the Robertson-Walker metric lacks singularities in the form of a horizon, we can confidently assume that our scenario is applicable to the interior of a black hole. Given that in the gravitational collapse $k$ has the dimension of
$M^2=L^{-2}$, the scale factor $R$ is dimensionless. To make a direct
contact with cosmology with $R$ of the dimension of $L$, we introduce the following rescaling into equation (\ref{LE})
\begin{equation}\label{rescaling}
a(t)=L_0 R(t),\quad
\widetilde{r}=\frac{r}{L_0},\quad L_0=\frac{1}{\sqrt{k}}.C\end{equation}
As $L_0$ carries the dimension of length, the scale factor $a$ now possesses the same dimension. This allows us to recast the line element (\ref{LE}) as
\begin{equation}\label{LE1}
ds^2=dt^2-a^2(t)\left[\frac{d\widetilde{r}^2}{1-\widetilde{r}^2}+\widetilde{r}^2 d\vartheta^2+\widetilde{r}^2\sin^2{\vartheta}d\varphi^2\right].
\end{equation}
This transformation provides us with a more straightforward avenue for drawing parallels between gravitational collapse and cosmology when $k=1$. Following the methodology outlined in \cite{KT,Inverno}, in order to remedy to the fact that the coefficient of $d\widetilde{r}^2$ becomes singular at $\widetilde{r}=1$, we circumvent this difficulty by introducing a new coordinate $\chi$, defined as $\widetilde{r}=\sin{\chi}$ with $0\leq\chi\leq\pi$. This relationship gives us $d\widetilde{r}=\cos{\chi}d\chi=\sqrt{1-\widetilde{r}^2}d\chi$ and makes it possible to rewrite the line element (\ref{LE1}) as
\begin{equation}\label{LE2}
ds^2=dt^2-a^2(t)\left[d\chi^2+\sin^2{\chi}\left(d\vartheta^2+\sin^2{\vartheta}d\varphi^2\right)\right].
\end{equation}
The spatial component of the metric (\ref{LE2}) describes a 3-dimensional surface, which we can position within a 4-dimensional Euclidean space specified by coordinates $(w, x, y, z)$. The relationships between these coordinates and our original ones are defined as follows \cite{Inverno}
\begin{equation}\label{trafos}
w=a\cos{\chi},\quad
x=a\sin{\chi}\sin{\vartheta}\cos{\varphi},\quad
y=a\sin{\chi}\sin{\vartheta}\sin{\varphi},\quad
z=a\sin{\chi}\cos{\vartheta}.
\end{equation}
The feasibility of such an embedding arises from the fact that the line element of the Euclidean metric, $d\sigma_E^2$, can be represented as
\begin{equation}
d\sigma_E^2 = dw^2+dx^2+dy^2+dz^2 =a^2(t)\left[d\chi^2+\sin^2{\chi}\left(d\vartheta^2+\sin^2{\vartheta}d\varphi^2\right)\right].
\end{equation}
Building further on equation (\ref{trafos}), we can deduce that
\begin{equation}
w^2+x^2+y^2+z^2=a^2(t),
\end{equation}
indicating that our 3-dimensional surface can be envisioned as a 3-dimensional sphere encapsulated within the 4-dimensional Euclidean space. To establish the WDW equation applicable to gravitational collapse of the manifold $\mathcal{M}$ described by (\ref{LE2}), we begin with an action consisting of that of Einstein gravity plus a possible cosmological term, $\Lambda$, and matter, given by \cite{Wilt,KT}
\begin{equation}\label{SEH}
S=-\frac{1}{16\pi G}\int_{\mathcal{M}} d^4 x\sqrt{-g}\left(R+2\Lambda\right)+\frac{1}{8\pi G}\int_{\partial\mathcal{M}}d^3 x\sqrt{-h} K+S_{matter},   
\end{equation}
where the Ricci scalar $R$ and the extrinsic curvature $K$ are \cite{KT}
\begin{equation}
R=-6\left[\frac{\ddot{a}}{a}+\left(\frac{\dot{a}}{a}\right)^2+\frac{1}{a^2}\right],\quad
K=-3\frac{\dot{a}}{a}.
\end{equation}
Here, $g$ represents the determinant of the metric corresponding to (\ref{LE2}) with $\sqrt{-g}=a^3(t)\sin^2{\chi}\sin{\vartheta}$ while $h$ is the determinant of the spatial part of the metric associated to (\ref{LE2}). Notice that in (\ref{SEH}), we included a Gibbons-Hawking-York term needed later to produce the correct equations of motions for manifolds with boundaries \cite{Gibbons1,York,Kaplan}. Let us start by noticing that the 3-volume of the spatial hypersurface is finite, namely
\begin{equation}
\int_{\partial\mathcal{M}} d^3 x\sqrt{-g}=\int_0^\pi d\chi\int_0^\pi d\vartheta\int_0^{2\pi}d\varphi\sqrt{-g}=2\pi^2 a^3(t)=\int_{\partial\mathcal{M}}d^3x\sqrt{-h}.
\end{equation}
This is however not the case in cosmologies with $k=0$ or $k=-1$ \cite{Inverno}. At this point, we can immediately integrate (\ref{SEH}) over the angular variables to obtain 
\begin{equation}\label{SHE1}
S=-\frac{3\pi}{4G}\int dt\left[-a^2\ddot{a}-a\dot{a}^2-a+\frac{\Lambda}{3}a^3\right]-\frac{3\pi}{4G}a^2\dot{a}+S_{matter}. 
\end{equation}
In order to get rid of the second derivative of $a$ in the expression above, we can use the following identity 
\begin{equation}
a^2\ddot{a}=\frac{d(\dot{a}a^2)}{dt}-2a\dot{a}^2
\end{equation}
in (\ref{SHE1}) combined with a straightforward integration. This leads to the result
\begin{eqnarray}
S&=&\frac{3\pi}{4G}a^2\dot{a}-\frac{3\pi}{4G}\int dt\left[a\dot{a}^2-a+\frac{\Lambda}{3}a^3\right]-\frac{3\pi}{4G}a^2\dot{a}+S_{matter},\\
&=&-\frac{3\pi}{4G}\int dt\left[a\dot{a}^2-a+\frac{\Lambda}{3}a^3\right]+S_{matter}.
\end{eqnarray}
The correct road to quantization pre-assumes a correct Lagrangian.  
We highlight here that constructing a matter Lagrangian that yields
the energy-momentum tensor of a perfect fluid is not a straightforward
enterprise \cite{Schutz,Ray} in general relativity. However, carefully
constraining the variation $\delta g_{\mu\nu}$, it is possible to find
a suitable candidate from the energy density $\rho$ which is a
scalar \cite{rho}. Indeed, such a choice guarantees that the
Euler-Lagrange equations give the full
Friedmann equations (no contraints are necessary here, see Appendix~\ref{ML}). Hence a suitable proposal is provided by
\begin{equation}
    S_{matter}=\int_{\mathcal{M}}d^4 x\sqrt{-g}L_{matter}=-\int_{\mathcal{M}}d^4 x\sqrt{-g}\rho=-2\pi^2\int a^3\rho dt,
\end{equation}
This leads us to the total Lagrangian of the form
\begin{equation}
L=-\frac{3\pi}{4G}\left(a\dot{a}^2-a+\frac{\Lambda}{3}a^3\right)-2\pi^2a^3\rho.
\end{equation}

\section{Setting up the WDW equation for the gravitational collapse}
Taking into account that the Hamiltonian $H=\pi_a\dot{a}-L$ has conjugate momentum \cite{KT} 
\begin{equation}
\pi_a=-\frac{2G}{3\pi}a\dot{a},    
\end{equation}
it is straightforward to check that
\begin{equation}\label{Hamiltonian}
H=-\frac{G}{3\pi}\frac{\pi_a^2}{a}+\frac{3\pi}{4G}\left(-a+\frac{\Lambda}{3}a^3\right)+2\pi^2 a^3\rho.
\end{equation}
Applying the canonical quantization prescription $\pi_a\to -id/da$, the  WDW equation $H\Psi(a)=0$ transforms into
\begin{equation}\label{SCH1}
\left[\frac{G}{3\pi a}\frac{d^2}{da^2}+\frac{3\pi}{4G}\left(-a+\frac{\Lambda}{3}a^3\right)+2\pi^2 a^3\rho\right]\Psi(a)=0. 
\end{equation}
It is important to note that we have chosen a factor ordering of
\begin{equation}\label{fa}
\pi^2_a\to-a^{-q}\left[\frac{d}{da}a^q\frac{d}{da}\right]    
\end{equation}
corresponding to $q=0$. This decision facilitates the transformation of equation (\ref{SCH1}) into the familiar form of a one-dimensional Schr\"{o}dinger equation for a particle with zero total energy and half the unit mass. Finally, we can rewrite (\ref{SCH1}) as
\begin{equation}\label{WDW01}
\left(-\frac{d^2}{da^2}+V_{eff}(a)\right)\Psi(a)=0,\quad
V_{eff}(a)=-\frac{9\pi^2}{4G^2}\left(-a^2+\frac{\Lambda}{3}a^4\right)-\frac{6\pi^3}{G}a^4\rho.
\end{equation}
At this stage, it is worth noting an important detail. A routine
examination of dimensions reveals that all terms within the round
brackets of equation (\ref{WDW01}) share a consistent dimension of
$M^2$. In contrast, a similar equation has been independently derived
in \cite{Braz} through different means, revealing a curious
discrepancy where not all terms maintain the same dimensionality. In the context of cold dust, where $\rho=\rho_0 (a_0/a)^3$, the corresponding WDW equation is 
\begin{equation}
\left[-\frac{d^2}{da^2}+\left(-\frac{3\pi^2\Lambda}{4G^2}a^4+\frac{9\pi^2}{4G^2}a^2-\frac{6\pi^3\rho_0 a_0^3}{G}a\right)\right]\Psi(a)=0.
\end{equation}
If we let $R\equiv \widetilde{a}=a/L_0$ with $L_0$ defined as in (\ref{rescaling}), the above equation can be rewritten as
\begin{equation}\label{WDWp}
\left[-\frac{d^2}{d\widetilde{a}^2}+\left(-\frac{3\pi^2\Lambda L_0^6}{4G^2}\widetilde{a}^4+\frac{9\pi^2 L_0^4}{4G^2}\widetilde{a}^2-\frac{6\pi^3\rho_0 \widetilde{a}_0^3 L_0^6}{G}\widetilde{a}\right)\right]\Psi(\widetilde{a})=0.
\end{equation}
Furthermore, if we normalise the radial coordinate $r$ so that $\widetilde{a}_0=1$ \cite{Weinberg}, then (\ref{WDWp}) becomes
\begin{equation}\label{WDWq}
\left[-\frac{d^2}{d\widetilde{a}^2}+\left(-\frac{3\pi^2\Lambda L_0^6}{4G^2}\widetilde{a}^4+\frac{9\pi^2 L_0^4}{4G^2}\widetilde{a}^2-\frac{6\pi^3\rho_0 L_0^6}{G}\widetilde{a}\right)\right]\Psi(\widetilde{a})=0.
\end{equation}
Finally, by introducing the Planck density, denoted as $\rho_{Pl}=1/G^2$, the vacuum density $\rho_{vac}=\Lambda/(8\pi G)$, and noting that $L_0^2=3/(8\pi G\rho_0)$, we can recast (\ref{WDWq}) into the final form
\begin{equation}\label{WDWfin}
\left(-\frac{d^2}{d\widetilde{a}^2}+U_{eff}(\widetilde{a})\right)\Psi(\widetilde{a})=0,
\end{equation}
where the effective potential is represented by the quartic polynomial
\begin{equation}\label{Veff}
U_{eff}(\widetilde{a})=\alpha \widetilde{a}^4+\beta\widetilde{a}(\widetilde{a}-1).
\end{equation}
The parameters in this equation are defined as
\begin{equation}\label{coeff}
\alpha=-\frac{\rho_{vac}}{\rho_0}\beta,\quad
\beta=\frac{81}{256}\left(\frac{\rho_{Pl}}{\rho_0}\right)^2.
\end{equation}
Transitioning to the de Sitter scenario, which is distinguished by a
positive cosmological constant ($\Lambda > 0$), implies a non-zero
vacuum energy density ($\rho_{vac} > 0$). To provide a grasp of the
magnitudes of the coefficients introduced in equation (\ref{coeff}),
let us consider a few illustrative densities. The Planck density,
$\rho_{Pl}$, is notably large at $5.1 \cdot 10^{96}$ Kg/m$^3$. In
stark contrast, the vacuum energy density, $\rho_{vac}$, is
significantly smaller, at $5.9 \cdot 10^{-27}$ Kg/m$^3$. For a typical
cloud core, we
can establish an order of magnitudes as follows for the forthcoming analysis
\begin{equation}\label{ordering}
\frac{\rho_{vac}}{\rho_0}\ll 1\ll\frac{\rho_{Pl}}{\rho_0},
\end{equation}
This implies that the terms with $\alpha$ in the effective potential
become influential only at large values of $\widetilde{a}$.

\section{Results}
\begin{figure}[ht!] 
    \includegraphics[width=0.3\textwidth]{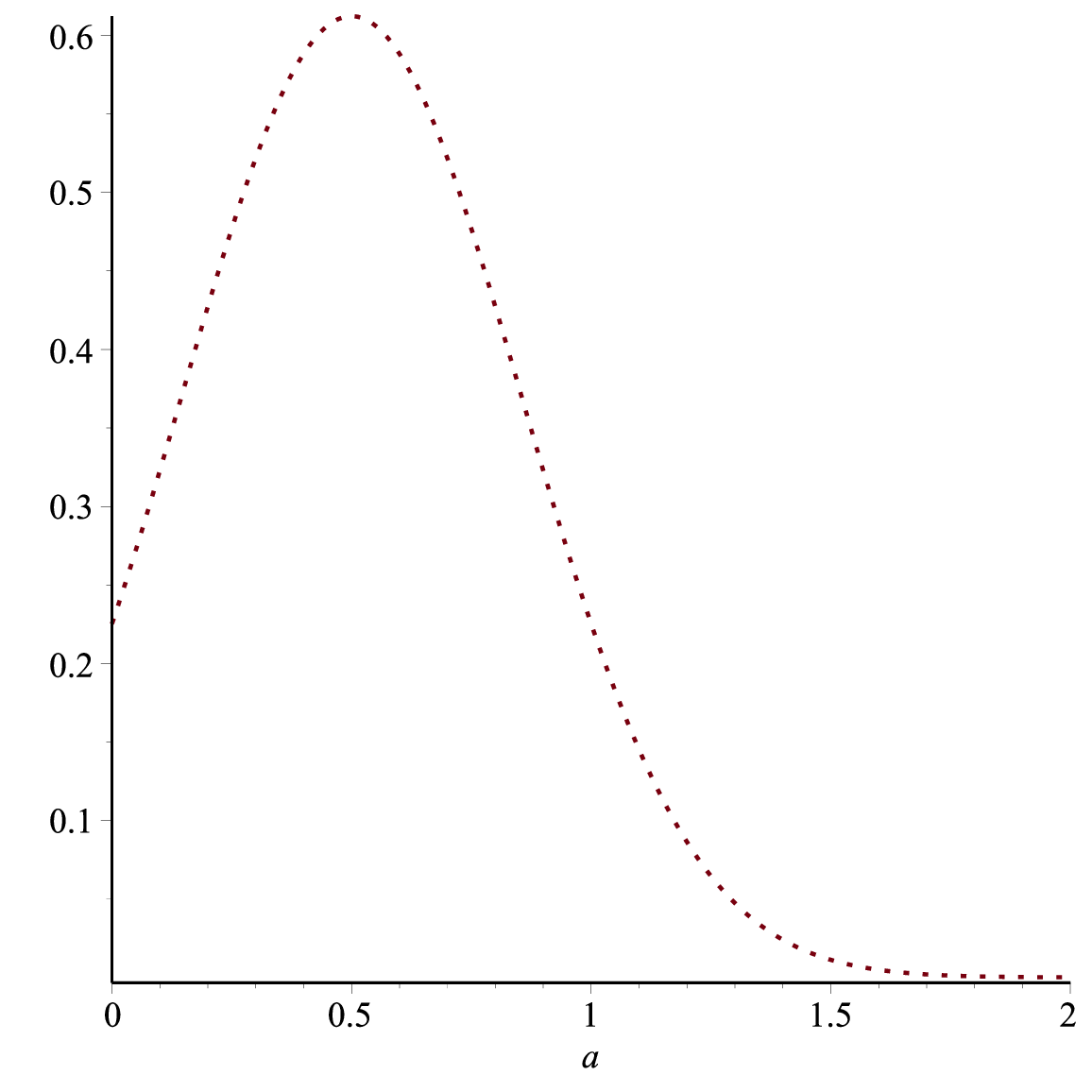}
    \includegraphics[width=0.3\textwidth]{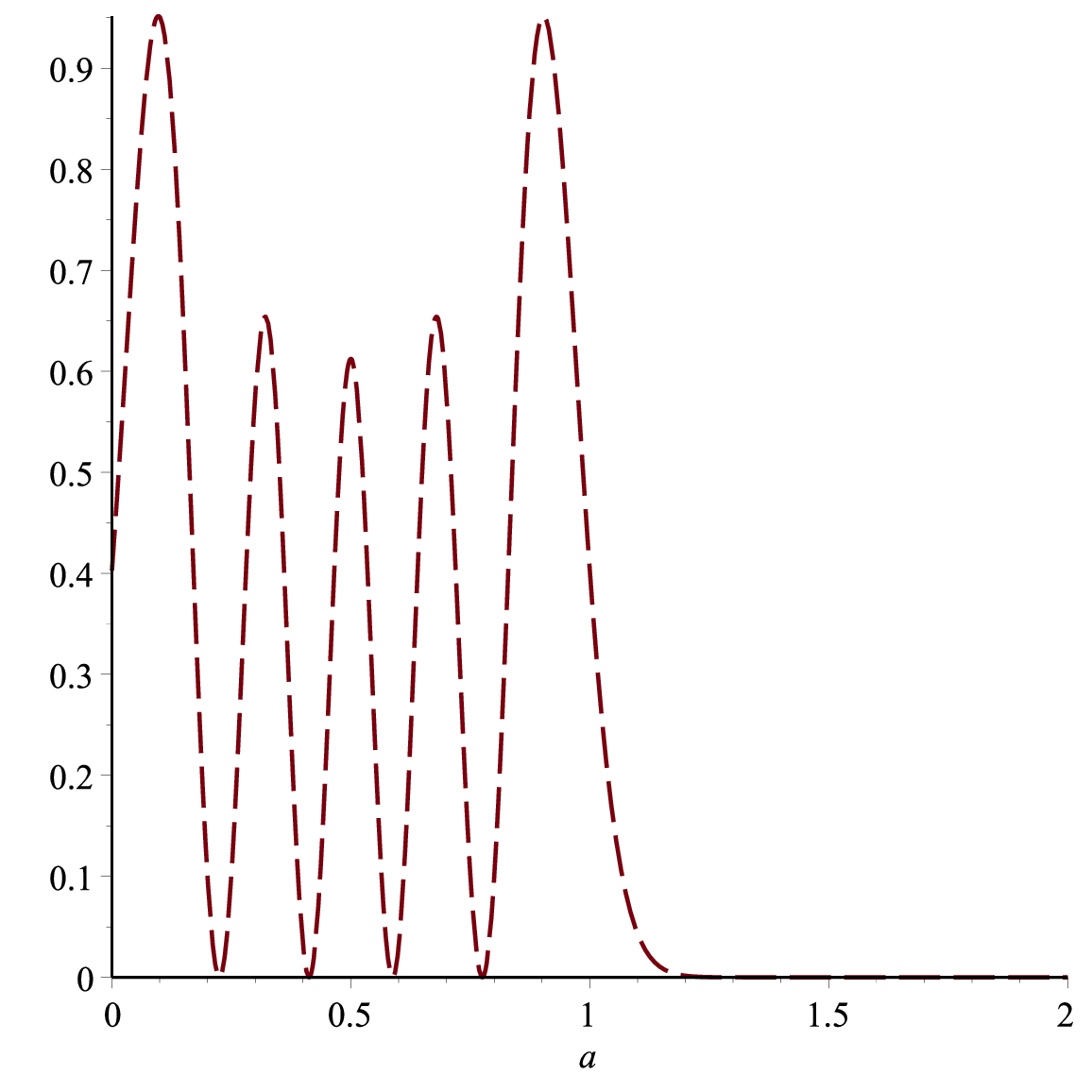}
    \includegraphics[width=0.3\textwidth]{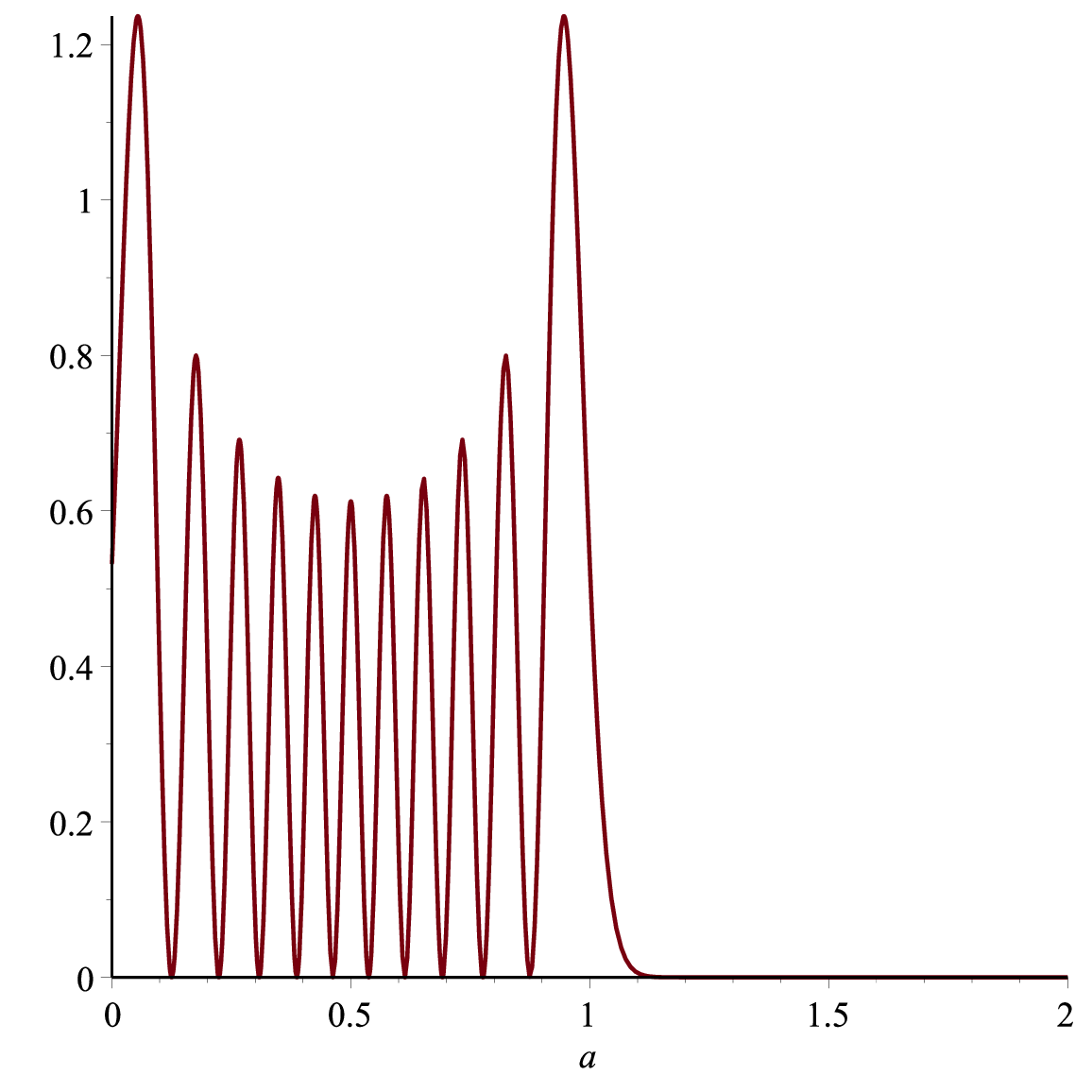}
\caption{\label{figure00}
Plot of the probability density $|\Psi_n|^2$ for the ground state $n=0$ (dotted line), $n=4$ (dashed line) and $n=10$ (solid line).
}
\end{figure}

We initially consider the case of $\Lambda=0$ which implies $\alpha=0$ in
the first approach. The effective potential simplifies to
\begin{equation}\label{VeffL}
U_{eff}(\widetilde{a})=\beta \widetilde{a}(\widetilde{a}-1).    
\end{equation}
and represents a parabola with a minimum at $\widetilde{a}=1/2$. Strikingly, by introducing the transformation $\widehat{a}=\widetilde{a}-1/2$, we can recast the WDW equation
\begin{equation}\label{WDW0}
-\frac{d^2\Psi}{d\widetilde{a}^2}+\beta \widetilde{a}(\widetilde{a}-1)\Psi(\widetilde{a})=0,
\end{equation}
subject to the boundary condition
\begin{equation}\label{BC1}
\lim_{\widetilde{a}\to+\infty}\Psi(\widetilde{a})=0    
\end{equation}
together with the normalization condition
\begin{equation}\label{normalization}
\int_0^\infty|\Psi(\widetilde{a})|^2~d\widetilde{a}=1.    
\end{equation}
into a form that mirrors the equation of a harmonic oscillator. Specifically, we can rewrite (\ref{WDW0}) as
\begin{equation}\label{Weber}
-\frac{d^2\Psi}{d\widehat{a}^2}+\beta\widehat{a}^2\Psi(\widehat{a})=\frac{\beta}{4}\Psi(\widehat{a}).
\end{equation}
In this reformulation, $\beta$ emerges as a characteristic parameter that quantifies the harmonic potential, thereby playing a fundamental role in the governing wave equation. If we introduce the dimensionless variable $\xi=\sqrt[4]{\beta}\widehat{a}$, (\ref{Weber}) becomes 
\begin{equation}\label{Herm}
\frac{d^2\Psi}{d\xi^2}=\left(\xi^2-K\right)\Psi(\xi),\quad
K=\frac{\sqrt{\beta}}{4}.
\end{equation}
which is reminscent of a dimensionless harmonic oscillator \cite{Grif}.
As a quick reminder on how to find a solution, we determine the permissible values of $K$ (and consequently, of $\beta$). First of all, we observe that for very large $\xi$, $\xi^2\gg K$ and in this regime $d^2\Psi/d\xi^2\sim \xi^2\Psi(\xi)$ which leads to the approximate solution $\Psi(\xi)\sim Ae^{-\xi^2/2}+Be^{\xi^2/2}$. Since the second term is not normalizable, we must pick $B=0$. This observation suggests the following ansatz
\begin{equation} \label{wave}
\Psi(\xi)=h(\xi)e^{-\xi^2/2},    
\end{equation}
which applied to (\ref{Herm}) leads to the Hermite differential equation
\begin{equation}
\frac{d^2 h}{d\xi^2}-2\xi\frac{dh}{d\xi}+(K-1)h(\xi)=0.    
\end{equation}
Following the methodology outlined in \cite{Grif}, we require the power series representation for $h$ to terminate, as this condition guarantees the existence of normalizable solutions. This requirement is met when $K=2n+1$, leading to
\begin{equation}\label{quant0}
\beta=16(2n+1)^2,\quad n=0,1,2,\cdots.
\end{equation}
Alternatively, employing equation (\ref{coeff}), this requirement can be reformulated as
\begin{equation}
\frac{\rho_{Pl}}{\rho_0}=\frac{64}{9}(2n+1),\quad n=0,1,2,\cdots,
\end{equation}
from which 
\begin{equation}\label{q0}
\rho_0=\frac{9\rho_{Pl}}{64(2n+1)},\quad n=0,1,2,\cdots.
\end{equation}
These conditions express the permissible ratios of Planck density to
the initial cloud core density that correspond to normalizable
solutions. Taking into account that $\rho_{Pl}=5.1 \cdot 10^{96}kg/m^3$, we
recover the classical value of $\rho_0$ when $n$ reaches the order of
magnitude of $10^{100}$.  Indeed, large $n$ gives us back the
classical picture. Switching back to the variable $\widetilde{a}$ and using (\ref{normalization}), it is not difficult to verify that the ground state wave function is 
\begin{equation}\label{zeroeig}
\psi_0(\widetilde{a})= c_0 e^{-2\left(\widetilde{a}-\frac{1}{2}\right)^2},\quad c_0=\frac{\sqrt{2}}{\sqrt{\sqrt{\pi}\left[1+\mbox{erf}(1)\right]}}.
\end{equation}
In the above expression, the normalization factor is calculated in accordance with equation (\ref{normalization}). For an arbitrary value of $n$, a detailed computation (see Appendix~\ref{normcn}) yields the subsequent formula for the normalization factor
\begin{equation}\label{cn}
c_n=\sqrt{\frac{2}{\sqrt{\pi}\left[n! 2^n(1+\mbox{erf}(\sqrt{2n+1}))+\frac{2}{\sqrt{\pi}}e^{-(2n+1)}\sum_{k=0}^{n-1}\left(\begin{array}{c}
n\\
k
\end{array}
\right)(-1)^{k+1}H^{(k)}_{n-1}(\sqrt{n+1})H_{n-k}(\sqrt{n+1})\right]}}. 
\end{equation}
In the above, $\mbox{erf}(\cdot)$ and $H_i(\cdot)$ denote the error function and the Hermite polynomial of degree $i$, respectively. Additionally, $H^{(k)}_i$ signifies the $k$-th derivative of the Hermite polynomial of degree $i$. Importantly, equation (\ref{cn}) accurately reproduces equation (\ref{zeroeig}) when $n=0$. Lastly, we provide plots of the probability densities for various $n$ values in Figure~\ref{figure00}.

In spite of the absence of the time parameter, we would speculate
  that natural order of events is from classical to quantum, i.e.
  from large $n$ to ground state.

\section{Remarks on the time problem in quantum gravity}
This issue touches upon a fundamental challenge within the entire
field of Quantum Gravity. In both classical physics and standard
quantum mechanics, time is treated as an external parameter, akin to
readings from a clock outside the system under observation. However,
in general relativity, time is intricately intertwined with the fabric of
the system itself. Consequently, in a quantum general relativistic
framework of a closed system, the conventional notion of time ceases
to exist. This discrepancy is at the heart of the 'problem of time' in
quantum gravity: a pervasive issue that extends beyond the WDW
equation, affecting various approaches including Loop Quantum Gravity
and well known in the literature \cite{time1, time2, time3, time4,
  time5, time5a, time6, time7, time8}. 
We include this small exposition on the time problem since it is
usually addressed in the context of quantum cosmology, but we deal
here with another closed system.

The quantum formulation of gravity, despite its detachment from
classical time evolution, is far from being redundant. It reveals
unique aspects that classical theories do not capture. For example,
when applying the WDW equation to cosmology, we
encounter intriguing scenarios like the 'universe from nothing' which
emerges from a tunneling perspective. In the realm of Loop Quantum
Gravity, this approach leads to the quantization of spatial properties
such as areas. In our article, we explore how these quantum gravity
frameworks can circumvent the issue of singularities and lead to the
quantization of physical quantities like density. Nonetheless, the
problem of time remains unresolved till today. What we have additionally
shown by applying the WDW equation to gravitational collapse, is that
the time problem usually discussed in the context of cosmology is
persistent all over quantum gravity.  There are various suggestions how to
  resolve the time problem out of which we just mention the notable
  concrete contributions from \cite{time9, time10}.
Each offers a unique perspective on integrating or redefining time within the quantum gravity context. 

For instance, \cite{time9} offers several key ideas concerning the
problem of time in quantum gravity, particularly in relation to the
WDW equation. It argues
that since the time parameter in the Schr\"{o}dinger equation is not
observable, it is consistent to assume that a closed system, like the
Universe, or black hole as we have shown, is in a stationary state. The dynamical evolution we observe can be described entirely in terms of stationary observables, dependent on internal clock readings rather than external coordinate time.
The article also emphasizes that observable changes in the world are
not dependent on external coordinate time. This is due to a
superselection rule similar to that for charge in quantum field
theory, implying that only operators commuting with the Hamiltonian
(and hence stationary) can be observables. Therefore, the observed
temporal behavior of a system is actually a dependence on some
internal clock time. Two more aspects are worth mentioning.
First, \cite{time9} illustrates how change is observed through
stationary observables, using the example of a system of spinning
particles. The observable time dependence is not on the external time
but on the relationships between dynamical variables, particularly
those representing clock readings. This concept aligns with how time
is measured by quantum clocks, as discussed by Peres, and emphasizes
that the dynamics of a system depend entirely on stationary
observables.

Secondly, it addresses whether it is necessary to have a law of
evolution if there is no observable difference between a stationary
and a nonstationary state of the Universe, or a black hole in our
  case. The authors in \cite{time9} show that the evolution of a system, as
  dictated by clock-time, can occur without any reference to the usual
  law of evolution but rather by correlations between the clock and
  the rest of the system. Finally, they conclude that the dynamics of a system depend upon internal clock time and not on coordinate time. This dependence is solely represented by stationary operators, which are the only observables in this context.

In \cite{time10}, the wave function of the universe is represented in a
manner that allows the scale factor to be considered as a time
variable, contributing to a dynamic picture of the universe despite
the time-independent nature of the wave function itself. Moreover, it suggests
utilizing internal geometrical or matter variables to define a physically
meaningful time. This approach allows the probability density to be
defined in relation to the scale factor or scalar field, providing a
dynamic interpretation of quantum cosmology. We believe that this approach is also
applicable to our scenario.

Finally, it is worth considering the question whether quantum
mechanics can provide a concept of time without direct reliance on the
parametric time variable, usually denoted as $t$. While this possibility seems feasible, its explicit implementation in the context of quantum gravity remains unclear. In this context, it is important to highlight two concepts. Firstly, there is the Salecker-Wigner-Peres clock
\cite{time11},  which measures the advancement of dynamics in discrete
steps. Secondly, we can highlight the concept of dwell time \cite{time12}. In one
dimension, it is defined as $\tau=\int dx \Psi^*\Psi/j$ where $j$ is the quantum mechanical conserved current. This concept can also be applied to the case of the WDW equation in connection with the Robertson-Walker metric.  Without the explicit dependence on $t$, we have $dj/da =0$, implying that $j$ is a constant. Typically, dwell time is used in tunneling phenomena, but there is no a priori eason why it cannot be applied more globally. Its relation with the scale factor can bex expressed as $d\tau/da=\Psi^*(a)\Psi(a)/j$. 

Each of these studies grapples with the previously mentioned issue of
time, a challenge we hope will be resolved in future research. In
passing, we notice that we get an insight into some aspects of 
physical processes that do not always need to involve time
\cite{time13}. For instance, we can study the geometry of a Keplerian
orbit without referring to time. This is the bare minimum that a
quantum gravity program will always deliver \cite{time14}.

\section{Conclusions and outlook}
This paper embarks on an exploration of the late gravitational
collapse within the framework of quantum gravity, employing the
    WDW  equation as the foundational tool. Our approach, grounded in
    the canonical quantization of general relativity, probes into the
    realm where the traditional concept of time is absent as in many
    other problems of quantum gravity. Our methodology involves formulating the WDW equation specific to scenarios of gravitational collapse. By interpreting the dependence of the wave function on configuration space variables, we manage to
   encapsulate the essence of dynamical change in a framework
   traditionally governed by the passage of time. This interpretation
   allows us to explore the quantum aspects of the late collapse of
   astrophysical systems, such as black holes, within the framework of the WDW equation. The wave function behavior ,in this context, reveals interesting details about the quantum dynamics involved in gravitational collapse. More precisely, the results obtained from our analysis provide new insights into the quantum behavior of collapsing systems. We observe that the formation of a central singularity is avoided and that the matter density is quantized in terms of the Planck density. Moreover, the original differential equation in the scale factor $\widetilde{a}$ is not exactly the standard harmonic quantum oscillator.  This becomes clear when looking at the argument $\xi$ of the wave function.  With
$\xi=\sqrt[4]{\beta}\widehat{a}$ the quantized version of $\beta$
enters the argument of the wave function. Secondly, our problem is
defined on the real line between zero and infinity which makes the
normalization factor quite different from the standard harmonic
oscillator. The same can be said about the expectation values which we address in this section. Apart from the quantization of the
density, the interpretation is a clear avoidance of the central
singularity. For higher $n$, the probability density $|\Psi_n|^2$ shows several peaks between zero and one which means that there are several preferred non-zero values for $\widetilde{a}$. At the same time the first left maximum gets shifted to the zero as we increase $n$.
Finally, for very large $n$ we get a continuum.  The spacetime is not discretized in a conventional geometric manner, but rather in a probabilistic way. The role of $\widetilde{a}=1$ becomes clear when we express the density as $\rho_0=M_0/(4\pi/3)r_s^3$, implying that we contain all the mass within the Schwarzschild radius $r_s$ (black hole). If we choose the dimensionful variable $a$, which may represent a characteristic length scale in our system, to coincide with the Schwarzschild radius $r_s$, i.e. $a=r_s$, then the corresponding dimensionless variable $\widetilde{a}$ becomes unity ($\widetilde{a}=1$). This normalization not only simplifies our analysis by setting a natural scale for the system under consideration but it also establishes the position of the horizon. Indeed, the bulk of the wave function  squared is in the interval $[0,
1]$ with a small {\it{leak}}  beyond $1$.  This renders the horizon fuzzy. It might also have to do with Hawking radiation, but it is difficult to describe a dynamical process in a formalism without time. If we loosely associate $n$ with time and its direction progresses from larger to smaller $n$,  this scenario would qualitatively align with black hole radiation. Here, the de-excitement from $n^{'}$ to $n<n^{'}$ would have a higher probability of being outside $1$ for the state $n$. In this picture, the ground state would be a black hole remnant. Of course, this aspect  remains to be explored in more detail in future undertakings.
\begin{figure}[ht!]
\centering
    \includegraphics[width=0.4\textwidth]{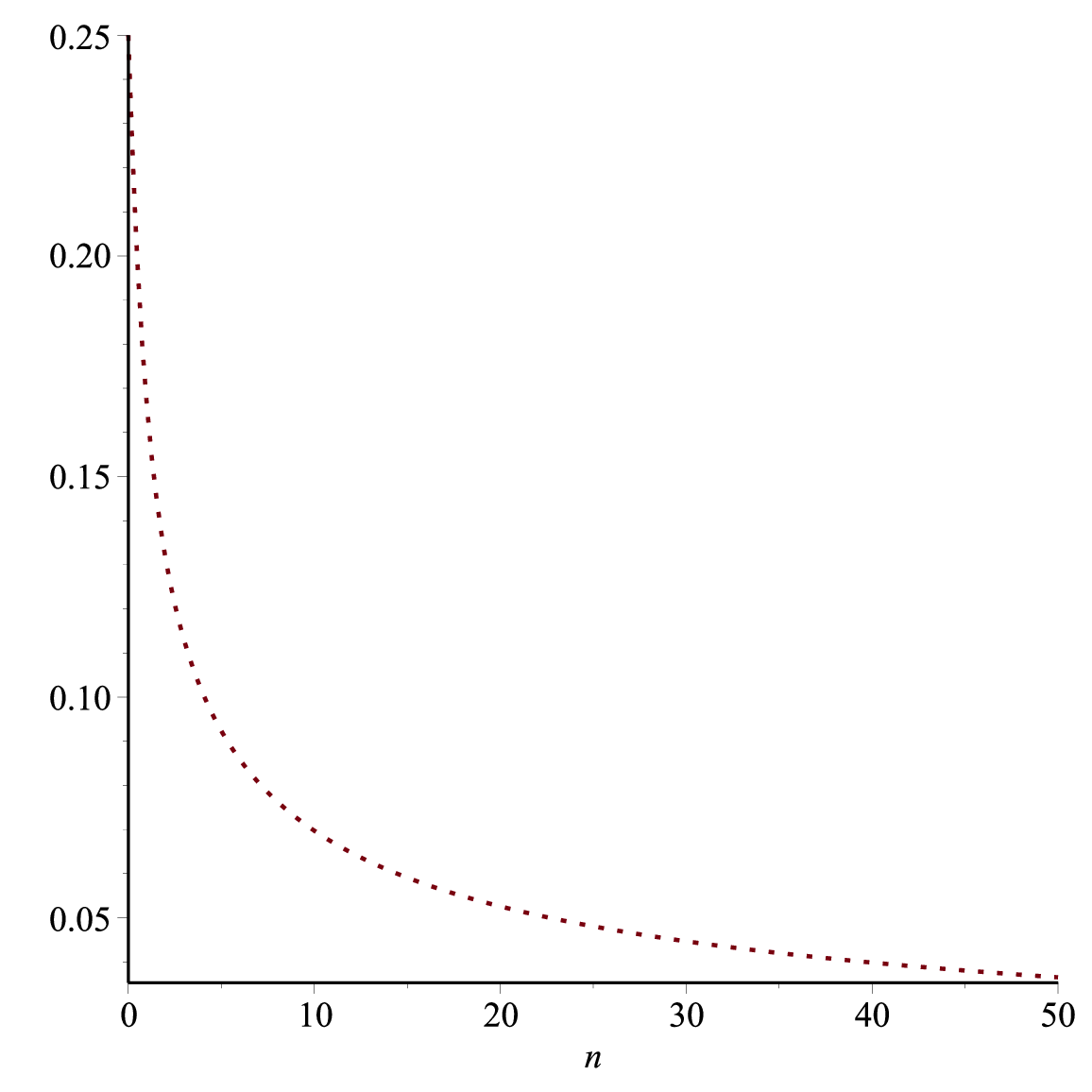}
\caption{
Plot of the expectation value $\langle\widetilde{a}\rangle$ for different values of $n$.
}
\label{figurexx}
\end{figure}
Finally, our conclusions about the preferred values of
$\widetilde{a}$ is confirmed by the
expectation value for $\widetilde{a}$, denoted here as
$\langle\widetilde{a}\rangle$. We can calculate it in accordance with the following formula
\begin{equation}\label{ev}
\langle\widetilde{a}\rangle=\int_0^\infty \psi^{*}_n(\widetilde{a})\widetilde{a}\psi_n(\widetilde{a})d\widetilde{a},\quad
\psi_n(\widetilde{a})=c_nH_n(\widetilde{a})e^{-2(2n+1)\left(\widetilde{a}-\frac{1}{2}\right)^2}.
\end{equation}
More specifically, our calculations reveal (see Appendix~\ref{44exp}) that
\begin{equation}\label{44}
\langle\widetilde{a}\rangle=f_1(n)+\frac{f_2(n)}{f_3(n)}
\end{equation}
with
\begin{align}
f_1(n)&=\frac{1}{4\sqrt{2n+1}},\\
f_2(n)&=e^{-(2n+1)}\left[2^{n-1}(n-1)n!+n\sum_{k=0}^n\left(\begin{array}{c}
n\\
k
\end{array}
\right)H^{(k)}_n(\sqrt{2n+1})H_{n-k}(\sqrt{2n+1})+\frac{1}{2}H^2_n(\sqrt{2n+1})\right],\\
f_3(n)&=2\sqrt{\pi}(2n+1)\left[n! 2^n\left(1+\mbox{erf}(\sqrt{2n+1})\right)+\frac{2e^{-(2n+1)}}{\sqrt{\pi}}\sum_{k=0}^{n-1}\left(\begin{array}{c}
n\\
k
\end{array}
\right)(-1)^{k+1}H^{(k)}_{n-1}(\sqrt{2n+1})H_{n-k}(\sqrt{2n+1})\right].
\end{align}
It is straightforward to confirm that $f_2/f_3$ asymptotically behaves
as $2^{-1}e^{-2n}$, thus indicating that $\langle\widetilde{a}\rangle$
approaches $0$ as $n\to\infty$. This behavior is also accurately
represented in Figure~\ref{figurexx}.  We remind the reader that in
the standard harmonic oscillator the expectation value of the position
$x$ is zero.

Last but not least, our work contributes to the dialogue on how
quantum mechanics and general relativity converge and interact,
particularly in extreme astrophysical conditions. We acknowledge certain limitations in our approach, particularly regarding the complexity of accurately modeling realistic astrophysical scenarios within the constraints of the WDW equation. The inherent assumptions and simplifications made to tackle the mathematical challenges also point to areas where further refinement is needed. Looking ahead, our study opens several directions for future research. One promising avanue is the exploration of more complex models of late gravitational collapse, incorporating additional factors such as the cosmological constant and different factor orderings. Another promising path is the development of numerical methods to solve the WDW equation for more realistic scenarios, which could provide a deeper understanding of the quantum aspects of gravitational collapse.

\appendix
\section{The matter Lagrangian}\label{ML}
 The Friedmann equations in the case of a perfect fluid are \cite{Carroll}
\begin{eqnarray}
\left(\frac{\dot{a}}{a}\right)^2+\frac{k}{a^2}&=&\frac{8\pi G}{3}\rho+\frac{\Lambda}{3},\label{uno}\\
\frac{\ddot{a}}{a}&=&-\frac{4\pi G}{3}\left(\rho+3p\right)+\frac{\Lambda}{3},\label{due}
\end{eqnarray}
where in the case of $\Lambda>0$ cosmological acceleration is possible. On the other hand, the equations which come directly from Einstein's field equations are (\ref{uno}) and
\begin{equation}\label{tre}
2\frac{\ddot{a}}{a}+\left(\frac{\dot{a}}{a}\right)^2+\frac{k}{a^2}=-8\pi G p+\Lambda.
\end{equation}
More precisely, (\ref{uno}) and (\ref{tre}) emerge from the $(0,0)$ and $(i,i)$ Einstein field equations, respectively \cite{Inverno}. Let us consider the following ansatz for the Lagrangian
\begin{equation}\label{Lagrange}
L=-\frac{3\pi}{4G}\left(a\dot{a}^2-ka+\frac{\Lambda}{3}a^3\right)-2\pi^2a^3\rho.    
\end{equation}
Given that $\pi_a=-(3\pi/2G)a\dot{a}$, the Euler-Lagrange equation $\dot{\pi}_a-\partial L/\partial a=0$ leads to
\begin{equation}
-\frac{3\pi}{2G}\left(\dot{a}^2+a\ddot{a}\right)+\frac{3\pi}{4G}\left(\dot{a}^2-k+\Lambda a^2\right)+ 2\pi^2\frac{d}{da}(a^3\rho)=0.
\end{equation}
Subsequently, multiplying the previous equation by $-(4G/3\pi)a^{-2}$ provides
\begin{equation}
2\frac{\ddot{a}}{a}+\left(\frac{\dot{a}}{a}\right)^2+\frac{k}{a^2}-\Lambda-\frac{8\pi G}{3}\frac{1}{a^2}\frac{d}{da}(a^3\rho)=0.
\end{equation}
Finally, we rearrange the terms to allow for easy comparison with (\ref{tre}) as follows
\begin{equation}\label{oben}
2\frac{\ddot{a}}{a}+\left(\frac{\dot{a}}{a}\right)^2+\frac{k}{a^2}=\frac{8\pi G}{3}\frac{1}{a^2}\frac{d}{da}(a^3\rho)+\Lambda.
\end{equation}
Evaluating the derivative in (\ref{oben}) results in
\begin{equation}\label{almostthere}
2\frac{\ddot{a}}{a}+\left(\frac{\dot{a}}{a}\right)^2+\frac{k}{a^2}= 8\pi G\left(\rho+\frac{a}{3}\frac{d\rho}{da}\right)+\Lambda
\end{equation}
At this point, it is worth mentioning that for a perfect fluid, we have the relations \cite{Carroll}
\begin{equation}\label{EOSCar}
p=w\rho,\quad\
\rho=ca^{-3(1+w)},  
\end{equation}
where $c$ is an integration constant. Therefore, the derivative of $\rho$ with respect to $a$ becomes
\begin{equation}
\frac{d\rho}{da}=-\frac{3}{a}(1+w)\rho.  
\end{equation}
This leads us to the conclusion
\begin{equation}\label{inter}
\rho+\frac{a}{3}\frac{d\rho}{da}=-w\rho.
\end{equation}
If we replace (\ref{inter}) into (\ref{almostthere}), we obtain
\begin{equation}
2\frac{\ddot{a}}{a}+\left(\frac{\dot{a}}{a}\right)^2+\frac{k}{a^2}=- 8\pi G w\rho+\Lambda.
\end{equation}
We can then express the above equation using the first equation in (\ref{EOSCar}) as
\begin{equation}
    2\frac{\ddot{a}}{a}+\left(\frac{\dot{a}}{a}\right)^2+\frac{k}{a^2}=- 8\pi G p+\Lambda.
\end{equation}
A direct comparison of this equation with (\ref{tre}) suggests that we should select the minus sign in (\ref{Lagrange}).

\section{Derivation of the normalization factor (\ref{cn})}\label{normcn}
Let us recall that Hermite polynomials can be obtained from the generating function as follows \cite{Way}
\begin{equation}\label{gf}
e^{2tx-t^2}=\sum_{n=0}^\infty H_n(x)\frac{t^n}{n!}\Rightarrow H_m(x)=\left.\frac{d^m}{dt^m}\right|_{t=0}e^{2tx-t^2}.    
\end{equation}
In order to compute the normalization factor (\ref{cn}), we need to evaluate the following integral
\begin{equation}
    I=\int_0^\infty \psi_n^*(\widetilde{a})\psi_n(\widetilde{a})d\widetilde{a}
\end{equation}
with $\psi_n$ given as in (\ref{ev}). Instead of considering the integral above, it turns out to be useful to introduce the following integral
\begin{equation}\label{Inm}
   I_{n,m}= \int_0^\infty \psi_n^*(\widetilde{a})\psi_m(\widetilde{a})d\widetilde{a}.
\end{equation}
Switching to the variable 
\begin{equation}
    \xi=2\sqrt{2n+1}\left(\widetilde{a}-\frac{1}{2}\right),
\end{equation}
letting $c=\sqrt{2n+1}$ and appying (\ref{gf}), the integral (\ref{Inm}) becomes
\begin{equation}
I_{n,m}=c_n c_m\int_{-c}^\infty e^{-\xi^2}H_n(\xi)H_m(\xi)d\xi
=c_n c_m\int_{-c}^\infty e^{-\xi^2}D^n_0\left(e^{2t\xi-t^2}\right)D^m_0\left(e^{2s\xi-s^2}\right)d\xi
\end{equation}
where by $D^n_0$ , we denote $n$-th derivative, evaluated at $t=0$. The latter can be rewritten as follows
\begin{equation}
    I_{n,m}=c_n c_m D_0^{n,m}\int_{-c}^\infty e^{-\xi^2+2(t+s)\xi-(s^2+t^2)}d\xi
\end{equation}
with $D_0^{n,m}$ representing the $n$- th derivative in $t$, $m$-th derivative in $s$, both evaluated at $t=s=0$. If we complete the square, we end up with
\begin{equation}
     I_{n,m}=c_n c_m D_0^{n,m}e^{2st}\int_{-c}^\infty e^{-\left[\xi-(t+s)\right]^2}d\xi.
\end{equation}
The integral over $\xi$ can be evaluate by means of the transformation $\omega=\xi-(t+s)$. Then, we have
\begin{equation}
     I_{n,m}=\frac{\sqrt{\pi}}{2}c_n c_m D_0^{n,m}e^{2st}\left[1+\mbox{erf}(t+s+c)\right].
\end{equation}
At this point, we can restrict our attention to the case $n=m$. Proceeding as in \cite{Schiff} we find that $D_0^{n,n}e^{2st}=2^n n!$. However, the computation of $D_0^{n,m}e^{2st}\mbox{erf}(t+s+c)$ is a bit more complicated. We first apply the formula for the $n$-th derivative of a product of two functions to get
\begin{equation}
D_0^{n,n}e^{2st}\mbox{erf}(t+s+c)
=\frac{d^n}{dt^n}\left[
\left.\sum_{k=0}^n
\left(\begin{array}{c}
n\\
k
\end{array}
\right)\frac{\partial^k}{\partial s^k}(e^{2st})\frac{\partial^{n-k}}{\partial s^{n-k}}\mbox{erf}(t+s+c)\right|_{s=0}\right]_{t=0}.
\end{equation}
Taking into account that \cite{Abra}
\begin{equation}\label{identities}
  \frac{\partial^k}{\partial s^k}(e^{2st})=(2t)^k e^{2st},\quad
  \frac{d^\ell}{dz^\ell}\mbox{erf}(z)=\frac{2}{\sqrt{\pi}}\frac{d^{\ell-1}}{dz^{\ell-1}}e^{-z^2}\quad\forall\ell\geq 1
\end{equation}
and shifting indices lead to
\begin{align}
 D_0^{n,n}e^{2st}\mbox{erf}(t+s+c)=&2^n\left.\frac{d^n}{dt^n}(t^n\mbox{erf}(t+s+c))\right|_{t=0}\label{B11}\\
 &+\frac{2}{\sqrt{\pi}}\frac{d^n}{dt^n}\left[
\left.\sum_{k=0}^{n-1}
2^{n-k-1}\left(\begin{array}{c}
n\\
n-k-1
\end{array}
\right)t^{n-k-1}e^{2st}\frac{\partial^{k}}{\partial s^{k}}e^{-(t+s+c)^2}\right|_{s=0}\right]_{t=0}.
\end{align}
Applying again the formula for the $n$-th derivative of a product of two functions to (\ref{B11}) together with the following representation of the Hermite polynomials \cite{Abra}
\begin{equation}\label{Hn}
    H_n(x)=(-1)^n e^{x^2}\frac{d^n}{dx^n}e^{-x^2}\quad\forall n=0,1,\cdots
\end{equation}
gives
\begin{equation}
   D_0^{n,n}e^{2st}\mbox{erf}(t+s+c)=2^n n!\mbox{erf}(c)+\frac{2}{\sqrt{\pi}}\sum_{k=0}^{n-1}(-1)^k 2^{n-k-1}\left(\begin{array}{c}
n\\
n-k-1
\end{array}
\right)\frac{d^n}{dt^n}\left[t^{n-k-1}H_k(t+c)e^{-(t+c)^2}\right]_{t=0}.
\end{equation}
In order to further simplify the above expression, we observe that at $t=0$
\begin{equation}
    t^{n-k-1}H_k(t+c)e^{-(t+c)^2}=t^{n-k-1}\left[H_k(c)e^{-c^2}+\mathcal{O}(t)\right].
\end{equation}
This suggests that when we apply $d^n/dt^n$ to the function above, only the case $k=n-1$ will contribute. Hence, we have
\begin{equation}
    D_0^{n,n}e^{2st}\mbox{erf}(t+s+c)=2^n n!\mbox{erf}(c)+\frac{2}{\sqrt{\pi}}(-1)^{n-1}\frac{d^n}{dt^n}\left[H_{n-1}(t+c)e^{-(t+c)^2}\right]_{t=0}.
\end{equation}
Finally, if we use again the formula for the $n$-th derivative of a product of two functions together with (\ref{Hn}), we obtain
\begin{equation}
D_0^{n,n}e^{2st}\mbox{erf}(t+s+c)=2^n n!\mbox{erf}(c)+\frac{2}{\sqrt{\pi}}e^{-c^2}\sum_{k=0}^{n-1}(-1)^{k+1} \left(\begin{array}{c}
n\\
k
\end{array}
\right)H^{(k)}_{n-1}(c)H_{n-k}(c),
\end{equation}
where $H^{(i)}_{j}$ denotes the $i$-th derivative of the Hermite polynomial of degree $j$. At this point, the normalization coefficient given in (\ref{cn}) can be easily extracted from the condition $I_{n,n}=1$.

\section{Derivation of the expectation value (\ref{44})}\label{44exp}
Starting from the definition
\begin{equation}
 \langle\widetilde{a}\rangle=\int_0^\infty \psi^{*}_n(\widetilde{a})\widetilde{a}\psi_n(\widetilde{a})d\widetilde{a},
\end{equation}
switching to the variable $\xi$ as in the previous section and taking into account the normalization condition give
\begin{eqnarray}\label{ient}
    \langle\widetilde{a}\rangle=\frac{1}{4\sqrt{2n+1}}+\frac{1}{4(2n+1)}\int_{-c}^\infty\xi\psi^2_n(\xi)d\xi. 
\end{eqnarray}
Let us introduce the integral
\begin{equation}
    \mathcal{I}_{n,m}=\int_{-c}^\infty\xi\psi_n(\xi)\psi_m(\xi)d\xi=c_n c_m D_0^{n,m}e^{2st}\int_{-c}^\infty\xi e^{-[\xi-(t+s)^2]}d\xi,
\end{equation}
where we used (\ref{ev}) and the same notation employed in the previous section. By means of the change of variable $\omega=\xi-(t+s)$, the integral above becomes
\begin{equation}
 \mathcal{I}_{n,m}=c_n c_m\left[D_0^{n,m}e^{2st}(t+s)\int_{-(t+s+c)}^\infty e^{-\omega^2}+D_0^{n,m}e^{2st}\int_{-(t+s+c)}^\infty \omega e^{-\omega^2}\right].
\end{equation}
Upon carrying out the integration, one arrives at
\begin{equation}\label{genint}
     \mathcal{I}_{n,m}=c_n c_m\left[\frac{\sqrt{\pi}}{2}D_0^{n,m}e^{2st}(t+s)+\frac{\sqrt{\pi}}{2}D_0^{n,m}e^{2st}(t+s)\mbox{erf}(t+s+c)+\frac{e^{c^2}}{2}D_0^n \left(e^{-(t+c)^2}\right)D_0^m \left(e^{-(s+c)^2}\right)\right].
\end{equation}
To compute the integral in (\ref{ient}), we consider the case $n=m$ in (\ref{genint}). First of all, a straightforward application of the formula for the $n$-th derivative of a product of two functions shows that
\begin{equation}
    D_0^{n,n}e^{2st}(t+s)=0.
\end{equation}
Moreover, we have
\begin{equation}\label{AA}
    D_0^{n,n}e^{2st}(t+s)\mbox{erf}(t+s+c)=\frac{d^n}{dt^n}\left[\sum_{k=0}^n
    \left(\begin{array}{c}
n\\
k
\end{array}
\right)\frac{\partial^k}{\partial s^k}\left(e^{2st}(t+s)\right)\frac{\partial^{n-k}}{\partial s^{n-k}}\left.\mbox{erf}(t+s+c)\right|_{s=0}
    \right]_{t=0}.
\end{equation}
On the other hand, for $0\leq k\leq n$
\begin{equation}
  \frac{\partial^k}{\partial s^k}\left(e^{2st}(t+s)\right)=k(2t)^{k-1},  
\end{equation}
which replaced into (\ref{AA}) leads to
\begin{align}
  D_0^{n,n}e^{2st}(t+s)\mbox{erf}(t+s+c)=&n2^{n-1}\frac{d^n}{dt^n}\left[t^{n-1}\mbox{erf}(t+s+c)\right]_{t=0}\label{C9}\\
  &+\frac{d^n}{dt^n}\left[\sum_{k=1}^{n-1}\left(\begin{array}{c}
n\\
n-k
\end{array}
\right)(n-k)(2t)^{n-k-1}\frac{\partial^k}{\partial s^k}\mbox{erf}\left.(t+s+c)\right|_{s=0}\right]_{t=0}.
\end{align}
Using the product formula in (\ref{C9}) together with (\ref{Hn})and the second identity in (\ref{identities}) yields
\begin{align}
   D_0^{n,n}e^{2st}(t+s)\mbox{erf}(t+s+c)=&\frac{2^n}{\sqrt{\pi}}(n-1)n! e^{-c^2}\\
   &+\frac{2}{\sqrt{\pi}}\sum_{k=1}^{n-1}\left(\begin{array}{c}
n\\
n-k
\end{array}
\right)(n-k)(-1)^{k+1}2^{n-k-1}\frac{d^n}{dt^n}\left[t^{n-k-1}H_{k-1}(t+c)e^{-(t+c)^2}\right]_{t=0}.
\end{align}
One more application of the product rule and of the identity (\ref{Hn}) gives the final result
\begin{equation}
    D_0^{n,n}e^{2st}(t+s)\mbox{erf}(t+s+c)=\frac{e^{-c^2}}{\sqrt{\pi}}\left[2^n (n-1)n! +2n\sum_{k=0}^n
    \left(\begin{array}{c}
n\\
k
\end{array}
\right)H^{(k)}_n(c)H_{n-k}(c).
    \right]
\end{equation}
Finally, a straightforward computation which makes use of (\ref{Hn}) shows that
\begin{equation}
D_0^n \left(e^{-(t+c)^2}\right)=(-1)^n e^{-c^2}H_n(c).
\end{equation}
By means of (\ref{genint}) and the above result, (\ref{ient}) becomes
\begin{equation}
\langle\widetilde{a}\rangle=\frac{1}{4\sqrt{2n+1}}+\frac{c^2_n e^{-c^2}}{4(2n+1)}
\left[
2^{n-1}(n-1)n!+n\sum_{k=0}^n
    \left(\begin{array}{c}
n\\
k
\end{array}
\right)H^{(k)}_n(c)H_{n-k}(c)+\frac{1}{2}H_n^2(c).
\right]   
\end{equation}
Replacing (\ref{cn}) into the above expression gives (\ref{44}).


\begin{thebibliography}{99}
\bibitem{Hawking}
S. Hawking,  {\it{Black hole explosions?}}, Nature {\bf{248}}, 30 (1974); S. Hawking,  {\it{Particle creation by black holes}}, Commun. Math. Phys. {\bf{43}}, 199 (1975). 

\bibitem{Unruh}
S. A. Fulling, {\it{Nonuniqueness of Canonical Field Quantization in Riemannian Space-Time}}, Phys. Rev. D {\bf{7}}, 2850 (1973); P. C. W. Davies, {\it{Scalar production in Schwarzschild and Rindler metrics}}, J. Phys. {\bf{8}}, 609 (1975); W. G. Unruh,  {\it{Notes on black-hole evaporation}}, Phys. Rev. D {\bf{14}}, 870 (1976).

\bibitem{Loop}
T. Thiemann, {\it{Loop Quantum Gravity: An Inside View}}, Lect. Notes Phys. {\bf{721}}, 185 (2007); C. Rovelli, {\it{Quantum Gravity}},  Cambridge Monographs on Mathematical Physics (2004); A. Ashtekar and E. Bianchi, {\it{A Short Review of Loop Quantum Gravity}}, Rep. Prog. Phys. {\bf{84}}, 042001 (2021).

\bibitem{Triangulation}
R. Loll, {\it{Discrete Approaches to Quantum Gravity in Four Dimensions}}, Living Rev. Rel. {\bf{1}}, {\bf{13}} (1998).

\bibitem{Causal}
D. P. Rideout and R. D. Sorkin, {\it{Classical sequential growth dynamics for causal sets}}, Phys. Rev. D. {\bf{61}}, 024002 (2000).

\bibitem{Path}
A. Di Tucci and J.-L. Lehners, {\it{The No-Boundary Proposal as a Path Integral with Robin Boundary Conditions}}, Phys. Rev. Lett. {\bf{122}}, 201302 (2019); S. Hawking,  {\it{Quantum cosmology}} in S. W. Hawking and W. Israel (Eds) {\it{Three Hundred Years of Gravitation}}, Cambridge University Press. pp. 631-651 (1989).

\bibitem{Noncommutative}
A. Connes, {\it{Noncommutative Geometry}},  Academic Press (1995); A. Connes, {\it{Noncommutative geometry and reality}}, J. Math. Phys. {\bf{36}}, 6194 (1995); P. Nicolini, {\it{Noncommutative Black Holes, The Final Appeal To Quantum Gravity: A
Review}}, Int. J. Mod. Phys. A {\bf{24}}, 1229 (2009).

\bibitem{String}
K. Becker, M. Becker and J. Schwarz, {\it{String theory and M-theory: A
modern introduction}}, Cambridge University Press (2007).

\bibitem{WdW}
B. S. DeWitt, {\it{Quantum Theory of Gravity. I. The Canonical Theory}}, Phys. Rev. {\bf{160}}, 1113 (1967); J. A. Wheeler, in Batelle Rencontres: {\it{1967 Lectures in Mathematics and Physics}}, edited by C. DeWitt and J. A. Wheeler (Benjamin, New York, 1968), p. 242.

\bibitem{Halliwell}
J. J. Halliwell, {\it{Derivation of the Wheeler-DeWitt Equation from a Path Integral for Minisuperspace Models}}, Phys. Rev. D {\bf{38}}, 2468  (1988).

\bibitem{Smolin1}
T. Jacobson and L. Smolin, {\it{Nonpertubative quantum geometries}}, Nucl. Phys. B {\bf{299}}, 295 (1988).

\bibitem{Smolin2}
C . Rovelli and L. Smolin, {\it{Loop space representation of quantum general relativity}},  Nucl. Phys. B {\bf{331}}, 80 (1990).

\bibitem{Mitschul}
H.-J. Mitschul,  {\it{Solutions to the Wheeler-DeWitt Constraint of Canonical Gravity Coupled to Scalar Matter Fields}},  Class. Quantum Grav. {\bf{10}}, L149 (1993).

\bibitem{Hartle}
J. B. Hartle and S. W. Hawking, {\it{Wave Function of the Universe}}, Phys. Rev. D {\bf{28}}, 2960 (1983).

\bibitem{Vilenkin}
A. Vilenkin, {\it{Approaches to quantum cosmology }}, Phys. Rev. D {\bf{50}}, 2581 (1994).

\bibitem{Page}
S. W. Hawking and D. N. Page, {\it{Operator ordering and the flatness of the universe}}, Nucl. Phys. B {\bf{264}}, 185 (1986).

\bibitem{Kont}
N. Kontoleon and D.L. Wiltshire, {\it{Operator ordering and consistency of the wave function of the universe}}, Phys. Rev. D {\bf{59}}, 063513 (1999).

\bibitem{Vil}
A. Vilenkin, {\it{Quantum cosmology and the initial state of the Universe}}, Phys. Rev. D {\bf{37}}, 888 (1988).

\bibitem{Braz}
H. S. Vieira and V. B. Bezerra, {\it{Class of solutions of the Wheeler-DeWitt equation in the Friedmann-Robertson-Walker universe}}, 	Phys. Rev. D {\bf{94}} 023511  (2016).

\bibitem{Gao}
D. He, D. Gao and Q.-Y. Cai, {\it{Dynamical interpretation of the wavefunction of the universe}}, Phys. Lett. B {\bf{748}}, 361 (2015).

\bibitem{Vieira}
H.S. Vieira, V.B. Bezerra, C.R. Muniz, M.S. Cunha and H.R. Christiansen, {\it{Some exact results on quantum relativistic cosmology: dynamical interpretation and tunneling phase}},  Phys. Lett. B {\bf{809}}, 135712 (2020).

\bibitem{He1}
D. He, D. Gao and Q.-Y. Cai, {\it{Spontaneous creation of the universe from nothing}}, Phys. Rev. D {\bf{89}}, 083510 (2014).

\bibitem{He2}
D. He and Q.-Y. Cai, {\it{Inflation of small true vacuum bubble by quantization of Einstein-Hilber action}}, Sci. China Phys. Mech. Astron. {\bf{58}}, 079801 (2015).

\bibitem{Steigl}
R. \v Steigl and F. Hinterleitner, {\it{Factor ordering in standard quantum cosmology}},  Class. Quantum Gravity {\bf{23}}, 3879 (2006).

\bibitem{Weinberg}
S. Weinberg, {\it{Gravitation and Cosmology: Principles and
    Applications of The General Theory of Relativity}}, John Wiley \&
Sons (1972).

\bibitem{Bambi}
C. Bambi, {\it{Regular Black Holes: Towards a New Paradigm of Gravitational Collapse}},  Springer Series in Astrophysics and Cosmology (2023); C. Bambi, {\it{Black Holes: A Laboratory for Testing Strong Gravity}}, Springer Nature (2017).

\bibitem{Malafarina}
D. Malafarina, {\it{Black Hole Bounces on the Road to Quantum Gravity}}, Universe {\bf{4}}, 92 (2018); D. Malafarina, {\it{Classical collapse to black holes and quantum bounces: A review}}, Universe {\bf{3}}, 48 (2017); H. Chakrabarty,
A. Abdujabbarov, D. Malafarina and C. Bambi, {\it{A toy model for a baby universe inside a black hole}}, Eur. Phys. J. C {\bf{80}}, 373 (2020); D. Malafarina, {\it{A Brief Review of Relativistic Gravitational Collapse}}, Astrophys. Space Sci. Libr. {\bf{440}}, 169 (2016).

\bibitem{time1}
C. J. Isham, {\it{Canonical Quantum Gravity and the Problem of Time}}, in
{\it{Integrable Systems, Quantum Groups, and Quantum Field Theories}}, Eds. L. A. Ibort and M. A. Rodriguez, Springer (1993).

\bibitem{time2}
K. V. Kuchar, {\it{Time and interpretations of quantum gravity}}, in Proceedings of the 4th Canadian Conference on General Relativity and Relativistic
Astrophysics , Eds. G. Kunstatter, D. Vincent, and J. Williams, World Scientific Publishing Company, Singapore (1992).

\bibitem{time3}
 C. Kiefer and P.Peter, {\it{Time in quantum cosmology}}, Universe 2022, {\bf 8}, 36 (2022).
 
\bibitem{Bode}
P. Bodenheimer, {\it{Stellar Structure and Evolution}}, Encyclopedia of Physical Science and Technology (Third Edition), Academic Press (2003).

\bibitem{Wilt}
D.L. Wiltshire, {\it{An introduction to quantum cosmology}}, arXiv:gr-qc/0101003 (2000).

\bibitem{KT}
E. W. Kolb and M. S. Turner, {\it{The Early Universe}}, Addison-Wesley (1989).

\bibitem{Inverno}
R. D'Inverno, {\it{Introducing Einstein's Relativity}}, Clarendon Press, Oxford (1998).

\bibitem{Gibbons1}
G. W. Gibbons and S. W. Hawking, {\it{Action Integrals and Partition Functions in Quantum Gravity}}, Phys. Rev. D {\bf{15}}, 2752 (1977).

\bibitem{York}
J. W. York, Jr, {\it{Role of conformal three geometry in the dynamics of gravitation}}, Phys. Rev. Lett. {\bf{28}}, 1082 (1972).

\bibitem{Kaplan}
D. E. Kaplan, T. Melia and S. Rajendran, {\it{The Classical Equations of Motion of Quantized Gauge Theories, Part I: General Relativity}}, arXiv:hep-th/2305.01798 (2023).

\bibitem{Schutz}
B. F. Schutz, {\it{Perfect Fluids in General Relativity: Velocity Potentials and 
a Variational Principle$^{*}$}}, Phys. Rev. D {\bf{2}}, 2762 (1970).

\bibitem{Ray}
J. R. Ray, {\it{Lagrangian density for Perfect Fluids in General Relativity}}, J. Math. Phys. {\bf{13}}, 1451 (1972).

\bibitem{rho}
J. D. Brown, {\it{Action functionals for relativistic perfect fluid}}, Class. Quantum Gravity {\bf 10}, 1579 (1993); O. Bertolami, F. S. N. Lobo and. J. Paramos, {\it{Nonminimal coupling of perfect fluids to curvature}}, Phys. Rev. D {\bf{78}}, 064036 (2008); S. Mendoza and S. Silva, {\it{The matter Lagrangian of an ideal fluid}}, Int. J. Geom. Methods Mod. Phys. {\bf{18}}, 2150059 (2021).

\bibitem{Carroll}
S. M. Carroll, {\it{Spacetime and Geometry}}, Addison Wesley (2004).

\bibitem{Grif}
D. J. Griffiths, {\it{Introduction to Quantum Mechanics}}, 2nd edition, Pearson Prentice Hall (2005).

\bibitem{Way}
C. E. Wayne and V. Zharnitsky, {\it{Exponential bound of the integral of Hermite functions product with Gaussian weight}}, J. Math. Anal. Appl. {\bf{517}}, 126544 (2023).
\bibitem{Schiff}
L. I. Schiff, {\it{Quantum Mechanics}}, McGraw Hill (1968).

\bibitem{time4}
J. B. Barbour, {\it{The timelessness of quantum gravity: I. The evidence
from the classical theory}}, Class. Quantum Grav. {\bf 11}, 2853 (1994);
J. B. Barbour, {\it{The timelessness of quantum gravity: II. The
appearance of dynamics in static configurations}}, (1994) {\bf 11}, 2875 (1994);
J. B. Barbour, {\it{The end of time: the next revolution in physics}}, Oxford University Press, Oxford (1999).

\bibitem{time5} 
J. B. Barbour, B. Z. Foster and N. O. Murchadha, {\it{Relativity without relativity}}, Class. Quantum Grav. {\bf 19}, 3217 (2002).

\bibitem{time5a} 
E. Y. Chua and C. Callender, {\it{No time for time from no-time}}, Philos. Sci. {\bf 88}, 1172 (2021).

\bibitem{time6}
S. B. Gryb, {\it{Jacobi's Principle and the Disappearance of Time}}, Phys. Rev. D {\bf 8}, 044035 (2010).

\bibitem{time7}
C. Rovelli, {\it{Quantum mechanics without time: A model}}, Phys. Rev. D {\bf{42}},  2638 (1990); C. Rovelli, {\it{Time in quantum gravity: An hypothesis}},  Phys.
Rev. D {\bf 43}, 442 (1991).

\bibitem{time8}
A. M. Frauca, {\it{Reassessing the problem of time of quantum gravity}}, Gen. Relativ. Gravit. {\bf 55}, 21 (2023).

\bibitem{time9}
D. N. Page and W. K. Wooters, {\it{Evolution without time: Dynamics
described by stationary variables}},  Phys. Rev. D {\bf{27}}, 2885 (1983).

\bibitem{time10}
A. Vilenkin, {\it{Boundary conditions in quantum cosmology}}, Phys. Rev. D {\bf{33}}, 3560 (1986).

\bibitem{time11}
H. Salecker and E.P. Wigner, {\it{Quantum Limitations of the Measurement of
Space-Time Distances}}, Phys. Rev. {\bf{109}}, 571  (1958); A. Peres, {\it{Measurement of time by quantum clocks}}, Am. J. Phys. {\bf48}, 552 (1980).

\bibitem{time12} 
M. B\"{u}ttiker nd R. Landauer, Phys. Rev. Lett. {\bf{49}}, 1739 (1982); N.G. Kelkar, H. M. Casta\~neda and M. Nowakowski, {\it{Quantum time scales
 in alpha tunneling}}, Europhys. Lett. 85, 20006 (2009).
 
\bibitem{time13}
L. Smolin, {\it{Time Reborn: From the Crisis in Physics to the Future of the Universe}},  Mariner Books (2014).

\bibitem{time14}
M. Di Mauro, S. Esposito and A. Naddeo, {\it{A road map for Feynman's adventures in the land of gravitation}}, Eur. Phys. J. H {\bf{46}}, 22 (2021).
 
\bibitem{Abra}
M. Abramowitz and I. A. Stegun, {\it{Handbook of Mathematical Functions with Formulas, Graphs, and Mathematical Tables}}, 9th printing, Dover: New York (1972).
\end{thebibliography}
\end{document}